\documentclass[epsfig,12pt]{article}
\usepackage{epsfig}
\usepackage{graphicx}
\usepackage{relsize}
\usepackage{amsmath,mathtools}
\usepackage{amsfonts}

\newcommand{\beq}{\begin{equation}}   
\newcommand{\eeq}{\end{equation}}
\newcommand{\beqn}{\begin{eqnarray}}   
\newcommand{\eeqn}{\end{eqnarray}}

\newcommand{\Z}{\mathbb{Z}}

\newcommand{\gsim}{\lower.7ex\hbox{$
\;\stackrel{\textstyle>}{\sim}\;$}}
\newcommand{\lsim}{\lower.7ex\hbox{$
\;\stackrel{\textstyle<}{\sim}\;$}}

\begin{document}

\begin{titlepage}
\begin{flushright}
FTPI-MINN-14/17, UMN-TH-3341/14, \\ IHES/P/14/21\\
\end{flushright}

\vspace{10mm}

\begin{center}
{  \bf{\large   Non-Abelian Moduli on Domain Walls}}
\end{center}


\begin{center}

 {\large 
 E. Kurianovych$^a$ and M. Shifman$^{a,b,c}$}
\end {center}

\begin{center}

$^{a}${\em School of Physics and Astronomy, 
 University of Minnesota,
Minneapolis, MN 55455, USA}

 \vspace{2mm}
 
$^b${\em William I. Fine Theoretical Physics Institute, University of Minnesota,
Minneapolis, MN 55455, USA}

 \vspace{2mm}
 
$^{c}${\it Institut des Hautes \'{E}tudes Scientifiques, 35 Route de Chartres, 91440 Bures-sur-Yvette, France}

\end {center}


\vspace{10mm}

\begin{center}
{\large\bf Abstract}
\end{center}

\hspace{0.3cm}
We study a simple model supporting domain walls
which possess
two orientational moduli in addition to the conventional translational modulus. 
This model is conceptually close to Witten's cosmic strings.  We observe an $O(3)$ sigma model on the wall world volume in the low-energy limit. We solve (numerically) classical 
static equations of motion and find 
the wall profile functions and the value of the coupling of the $O(3)$ model in terms of the bulk parameters. 
In the second part a term describing a spin-orbit interaction in the bulk is added, which gives rise to an entanglement between rotational and translational moduli.  The corresponding extra term in the low-energy Lagrangian is calculated.
	
\vspace{2cm}

\end{titlepage}

\newpage

\section{Introduction}

Some field theories support topological defects with non-Abelian moduli on the world volume of these defects.
This discovery was first made in 2003 in the context of supersymmetric Yang-Mills theories with matter (for a 
thorough review and an extensive list of references see \cite{book}). Since then much effort has been invested 
in the studies of the so-called non-Abelian strings --  topological defects of codimension two. In this 
paper we develop a similar construction of codimenion one, i.e. non-Abelian walls. To this end we generalize
the analysis of the simplest model \cite{Sh1} supporting topologically stable strings with 
non-Abelian moduli.\footnote{In \cite{Sh1} the basic idea of domain walls with non-Abelian moduli due to the ``$\chi$-structure" (see below)
was outlined,
but calculations of the wall profiles were not carried out.} The model \cite{Sh1} was inspired
by Witten's construction for cosmic strings \cite{Wit} and used in \cite{ShYu,NShV,MShYu} 
to study diverse effects associated with non-Abelian moduli on the string world sheet.  

Our task is to identify non-Abelian moduli on 
domain walls and study their low-energy dynamics. We will show that, as in   \cite{Sh1}, the latter will be 
described by the O(3) sigma model.  Then we deform our basic model by adding a Lorentz violating term of a 
special form (relevant to condense-matter systems)  in the bulk. This term (similar to that 
in  \cite{NShV}) leads to a coupling between translational and orientational moduli. 

The qualitative non-Abelian string  picture  \cite{Sh1} is supported in the wall case by numerical calculations, 
in a way similar to that in \cite{MShYu} for the Abrikosov-Nielsen-Olesen string. In the wall case the emphasis 
is put on qualitative aspects too, rather than on exact results. Calculations are carried out in the quasiclassical 
approximation at weak coupling.\footnote{For technical reasons the relevant coupling constants, although small, 
are not too small. We are at the borderline of the quasiclassical domain.}

We begin with a simple domain wall  in a model of   a single real scalar field, and then add an extra term 
with a $\chi^i$ field in the Lagrangian. It is shown that the standard domain wall solution becomes unstable.
A nonvanishing $\chi^i$ develops in the wall core resulting in rotational moduli. For a fixed set of parameters 
the two-field solution is found numerically from the standard Euler-Lagrange equations. We discuss both translational 
and rotational moduli. In the last part a spin-orbit interaction is introduced. It gives rise to an entanglement of 
translational and rotational moduli.

\section{Domain wall and an additional field}

Let us start from a simple model (e.g. \cite{ShBook}, section 5) of a single real scalar field, described by the Lagrangian 
\beq
{\cal L}_0=\frac{1}{2}\partial_{\mu}\varphi\partial^{\mu}\varphi-V(\varphi), 
\qquad
V(\varphi)=\lambda(\varphi^{2}-v^{2})^{2},
\label{L0}
\eeq
where $\lambda$ and $v$ are constants.  This Lagrangian obviously possesses 
a $\Z_2$ symmetry ($\varphi\to  -\varphi$) which is spontaneously broken in the vacuum. Hence, the model 
supports domain walls. In this model there are two physically equivalent vacua. For a wall, oriented 
in the $(x,y)$ plane, we have the boundary condition $\varphi=\pm v$ for $z=\mp \infty$ (or vice versa). 
In this case the solution for $\varphi$ can be found analytically,
\beq
\varphi(z)=-v\tanh\left[\frac{m_{\varphi}}{2}(z-z_{0})\right],
\label{tanh}
\eeq
where $m_{\varphi}=\sqrt{8\lambda v^{2}}$ is the mass of the $\varphi$  field, and $z_{0}$ is the wall center. 
The minus sign in the right-hand side of (\ref{tanh})  was chosen for convenience of the subsequent consideration. 

Now let us add, according to \cite{Sh1}, a triplet field $\chi^{i}$, $i=1,2,3$, described by the Lagrangian
\beqn
{\cal L}_{\chi}&=&\frac{1}{2}\partial_{\mu}\chi^{i}\partial^{\mu}\chi^{i}-U(\chi,\varphi),
\label{Lchi}
\nonumber\\[2mm]
U(\varphi,\chi^i)&=&\gamma\left[(\varphi^{2}-\mu^{2})\chi^{i}\chi^{i}
+\beta(\chi^{i}\chi^{i})^{2}\right], \qquad v^2>\mu^2.
\label{U}
\eeqn
The deformed model with the extra field $\chi^i$ is described by the Lagrangian
\beq
{\cal L}={\cal L}_0+{\cal L}_{\chi}.
\label{Ltot}
\eeq

In the vacuum, where $\varphi=\pm v$, the expectation value of $\chi$ is $0$, but in the core of the wall, 
where $\varphi$ is small, $\chi$ develops a non-vanishing expectation value
\beq
\chi_{0}\cong\sqrt{\frac{\mu^2}{2\beta}}\, .
\label{chi0}
\eeq

\section{Instability of the \boldmath{$\chi=0$} solution}

The stability analysis is similar to that in \cite{Wit} and \cite{MShYu}. Let us assume that the 
field $\varphi$ is given by  (\ref{tanh}) and $\chi$ is small. This allows us to work in the quadratic 
in $\chi$ approximation. We have the following energy functional
\beq
{\cal E}_\chi=A\int\left[\frac{1}{2}\left(\frac{\partial\chi}{\partial z}\right)^2+\gamma(\varphi^2-\mu^2)\chi^2\right]dz,
\eeq
where $A$ is the area of the wall. After integration by parts we arrive at
\beq
{\cal E}_\chi=A\int\left[-\frac{1}{2}\chi\frac{\partial^2\chi}{\partial z^2}+\gamma(\varphi^2-\mu^2)\chi^2\right]dz\,.
\label{7}
\eeq
Whether or not $\chi=0$ is a stable solution depends on the eigenmodes of the operator in (\ref{7}). 
Existence of negative mode(s) will imply instability.

The eigenvalue equation can be interpreted as a Schr\"{o}dinger equation. Taking $\varphi$ 
from (\ref{tanh}) directly, we have the eigenvalue equation 
\beq
-\chi''+\frac{\gamma}{4\lambda}{\cal V}(\zeta)\chi=\epsilon\,\chi\,,
\label{Shrod}
\eeq
where
\beq
{\cal V}(\zeta)=\left[\left(1-\frac{\mu^2}{v^2}\right)-\frac{1}{\cosh^2 \frac{\zeta}{2}}\right]\,.
\label{ShrodPot}
\eeq
Moreover, $\zeta=m_\varphi z$, differentiation over $\zeta$ is denoted by prime, and the center of the wall is $\zeta_0=0$. 

\begin{figure}[t]
      \epsfxsize=250px
   \centerline{\epsffile{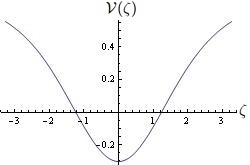}}
   \caption{Potential ${\cal V}(\zeta)$ in  the Schr\"{o}dinger equation (\ref{Shrod}).}
   \label{Coshplot}
\end{figure}

A plot of the potential (\ref{ShrodPot}) for the set of parameters used (see Sec. \ref{secf} 
for our numerical choice) is depicted in Fig. \ref{Coshplot}. We see a negative-energy domain near the origin, 
which allows in principle the existence of negative modes. Concretely, the lowest-eigenenergy $\epsilon$ was 
found numerically to be $-0.11$ for the choice of parameters presented in (\ref{parameters}).

\section{Equations and solutions}
\label{secf}

The Lagrangian (\ref{Ltot}) implies the following classical static equations of motion:
\beq
\begin{dcases}
\varphi''=4\lambda\varphi(\varphi^{2}-v^{2})+2\gamma\chi^{2}\varphi,  \\[2mm]
\chi^{i}~''=2\gamma(\varphi^{2}-\mu^{2})\chi^{i}+4\beta\gamma\chi^{2}\chi^{i},
\end{dcases}
\eeq
where $\chi^{2}=\chi^{j}\chi^{j}$ $(i,j=1,2,3)$, and $z$ differentiation is denoted by prime. The wall ansatz is implied:
the solution sought for is independent of $x$ and $y$.

The function $\chi(z)$ is determined by energy minimization, but  orientation of $\chi^i$ in the internal 
space can be arbitrary. Thus it is obvious that we can use an ansatz with a fixed orientation, for example,
\beq
\chi^{i}=\chi(z)\begin{pmatrix}0\\
0\\
1
\end{pmatrix}.
\eeq
This gives us
\beq
\begin{dcases}
\varphi''=4\lambda\varphi(\varphi^{2}-v^{2})+2\gamma\chi^{2}\varphi ,  \\[2mm]
\chi''=2\gamma(\varphi^{2}-\mu^{2})\chi+4\beta\gamma\chi^{3}\,.
\end{dcases}
\label{syst}
\eeq
We  see that this system of equations has the first integral
\beq
\frac{1}{2}\left(\frac{d\varphi}{dz}\right)^{2}+\frac{1}{2}\left(\frac{d\chi}{dz}\right)^{2}-W(z)=0,
\label{firstint}
\eeq
where 
\begin{figure}[t]
   \epsfxsize=324px
   \centerline{\epsffile{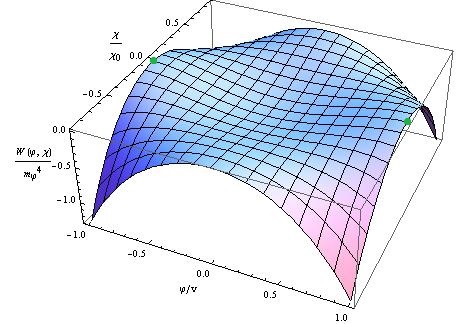}}
   \caption{Potential $-W(z)$ as a function of $\varphi$ and $\chi$. The points denoted 
   by $\bullet$ ($\varphi=\pm v$ and $\chi=0$) correspond to the vacua of the system, and a 
   domain wall can be visualized as a trajectory of motion in this potential with time standing for the $z$ coordinate.}
   \label{Potentialplot}
\end{figure}
\beq
W(z)=U\left[\varphi(z),\chi(z)\right]+V[\varphi(z),\chi(z)],
\eeq 
with $V$ from (\ref{L0}) and $U$ from (\ref{U}). The constant was chosen to vanish, 
because at infinity $\varphi=v$, $\chi=0$, and the kinetic term is zero. 
Equation (\ref{firstint}) allows us to visualize Eq.(\ref{syst}) as equations of 
motion for a particle in a two-dimensional potential $-W(z)$, where $\varphi$ and $\chi$ 
play the role of coordinates, and $z$ the role of time.

The ``potential" $-W(z)$ as a function of $\varphi$ and $\chi$ is plotted in Figure \ref{Potentialplot}. Parameters used are
\beq
\lambda=\frac{1}{12}\,, \quad \beta=0.2\,, \quad\gamma=\frac{2}{3}\,, \,\,\quad \mathlarger{\frac{\mu}{v}}=0.55\,.
\label{parameters}
\eeq
We  use this set of parameters for finding a sample solution. The quartic constants are moderately small which
warrants the validity of the quasiclasssical approximation. 

The unstable maxima of $-W(z)$ with $\varphi=\pm v$ and $\chi=0$ correspond to the vacua of the system, and  
trajectories connecting these points correspond to the domain wall configurations. We have a 
nonvanishing solution for $\chi$ in the wall core when the ``particle" is ``dragged" near the 
starting point in the $\chi$ direction, and then goes along $\varphi$. To have a wall without 
$\chi$, the ``particle" should move exactly in the $\varphi$ direction, but, as it was proved 
previously, for a certain set of parameters such a configuration is unstable.

Before passing to actual calculations it is worth asking what happens  if we vary the value of the parameters in (15), in particular, $\gamma$. When $\gamma$  is very small, only the $\phi$  field in the solution does not vanish. With increasing $\gamma$ metastable vacua in which $\chi$ condenses while $\phi$  does not start to ``attract” the wall until it becomes unstable,  and a new configuration, with a nonvanishing $\chi$ in the middle  appears. 

With further increase of $\gamma$, at a certain value of $\gamma$, $\phi$ and $\chi$  vacua will become degenerate, but this does not affect the (in)stability of the wall configuration. This is clear from Eq. (8) in which the potential ${\mathcal V}$ will still have negative modes. If we further increase $\gamma$ the $\phi$ vacua will become metastable. 

The choice of parameters in (15) was determined by the condition that $\gamma$ should be large enough to create a stable wall configuration with non-zero $\chi$ and, at the same time, small enough not to make $\phi$ vacua metastable. The choice we made is quite close to the limit of degenerate vacua. This particular choice was made only for the sake of calculational convenience. Generally speaking,  there is no need to fine-tune $\gamma$  to be close to this limit.

Now let us proceed to numerical solution of the system (\ref{syst}). Since we have boundary conditions set at infinity, 
we need to examine the asymptotic behavior of our equations there.

If $\varphi(-\infty)=v$, $\chi(-\infty)=0$, then at $z$ near minus infinity  $\chi$ and $\eta\equiv v-\varphi$ are small; 
therefore,
\beq
\begin{dcases}
\eta''=8v^{2}\lambda\eta,  \\[2mm]
\chi''=2\gamma(v^{2}-\mu^{2})\chi,
\end{dcases}
\eeq
which implies the following asymptotic behavior:
\beq
\eta=A\exp\left(\sqrt{8v^{2}\lambda}~z\right),
\qquad
\chi=B\exp\left[\sqrt{2\gamma(v^{2}-\mu^{2})}~z\right]
\eeq
at $z\rightarrow-\infty$. Here $A$ and $B$ are some unknown constants.

\begin{figure}[t]
   \epsfxsize=250px
   \centerline{\epsffile{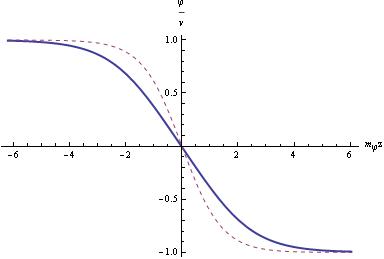}}
   \caption{The solid line represents  $\varphi(z)$ in the presence of $\chi$, the dashed line is $\varphi(z)$ for $\chi=0$, given by (\ref{tanh}).}
   \label{Phi}
\end{figure}

\begin{figure}
   \epsfxsize=250px
   \centerline{\epsffile{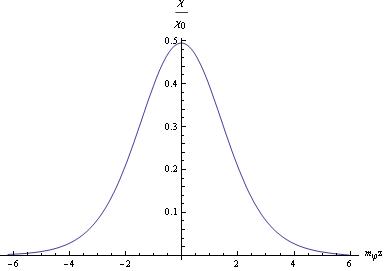}}
   \caption{Calculated $\chi(z)$.}
   \label{Chi}
\end{figure}

For a system of the second-order equations we need boundary conditions
for both functions themselves and their first derivatives. Asymptotical solutions
above provide us with the connection between functions and their derivatives
at infinity. Since we need both $\varphi$ and $\chi$ at the boundary and the analytical connection between them is 
unknown, the shooting method was used: we
set a small initial value for $\eta$ at $z=-\infty$ (on the left edge of the interval under consideration) and 
varied the initial value of $\chi$
in solving the system of equations, until we obtained the desired value of $\varphi$
at $+\infty$ (the right edge of the interval under consideration). The results obtained
are presented in Figs. \ref{Phi} and \ref{Chi}, where $\chi$ is normalized 
by $\mathlarger{\chi_{0}=\frac{\mu}{\sqrt{2\beta}}}$ and $z$ normalized by $m_\varphi$. 
For comparison the plot of $\varphi$ from (\ref{tanh}) (i.e. with $\chi=0$) is also given in Fig. \ref{Phi}.

\section{Moduli}

If we have the solutions for $\varphi(z)$ and $\chi(z)$, then due to
rotations in the internal space the general solution has rotational moduli 
\beq
\chi^{i}=\chi(z)\cdot S^{i}(x,y,t), \qquad S^{i}S^{i}=1, \hspace{3 px} i=1,2,3\,.
\eeq
The kinetic term takes the form
\beq
\partial_{\mu}\chi^{i}\partial^{\mu}\chi^{i}\rightarrow\left[
\chi(z)^{2}\right]\partial_{p}S^{i}\partial^{p}S^{i}-\left(\frac{d\chi}{dz}\right)^{2},\qquad p=0,1,2\,,
\eeq
which leads us to the following term in action on the world volume
\beq
{\Delta}S_{1}=\frac{1}{2}\,\frac{\chi_{0}^{2}}{m_{\varphi}}I_{1}\int dt\,dx\,dy\,\left(\partial_{p}S^{i}\partial^{p}S^{i}\right). 
\label{modact}
\eeq
Here
\beq
I_{1}=\frac{m_{\varphi}}{\chi_{0}^{2}}\int\chi(z)^{2}dz.
\eeq
From numerical integration over $z$, using the known function $\chi(z)$, we obtain 
\beq
I_{1}=0.38\,.
\eeq

Next, we should take into account the fact that
small excitations  due to the translational symmetry 
breaking on the wall, give rise to a single translational modulus. 
Under translations the function $\chi(z)$ does not change but the center of the wall 
is displaced, $$\chi(z)=\chi[z-z_{0}(t,x,y)]\,.$$
As a result,
\beq
\partial_{\mu}\chi^{i}\partial^{\mu}\chi^{i}\rightarrow\left(\frac{d\chi}{dz}\right)^{2}
\left[\partial_{p}z_{0}\partial^{p}z_{0}-1\rule{0mm}{4mm}\right],
\label{tranmodkinterm1}
\eeq
implying the emergence of an additional term in the action, 
\beq
{\Delta}S_{2}=\frac{1}{2}\,\chi_{0}^{2}m_{\varphi}I_{2}\int dt\,dx\,dy \,
\left(\partial_{p}z_{0}\partial^{p}z_{0}\right),
\label{tranmodact1}
\eeq
where
\beq
I_{2}=\frac{1}{\chi_{0}^{2}m_{\varphi}}\int\left(\frac{d\chi}{dz}\right)^{2}dz.
\eeq
From numerical integration we obtain $I_{2}=0.14$.

Let us add this term to the world-volume action for the field $\varphi$,
\beq
\frac{1}{2}\left[\int\left(\frac{d\varphi}{dz}\right)^{2}dz\right]\int dt\,dx\,dy \,
\left(\partial_{p}z_{0}\partial^{p}z_{0}\right).
\label{phimodaction}
\eeq
Equation (\ref{firstint}) imply  for the tension  
\beq
T=\int\left[\frac{1}{2}\left(\frac{d\varphi}{dz}\right)^{2}
+\frac{1}{2}\left(\frac{d\chi}{dz}\right)^{2}+V(z)\right]dz=\int\left[\left(\frac{d\varphi}{dz}\right)^{2}
+\left(\frac{d\chi}{dz}\right)^{2}\right]dz.
\label{tensionandderivatives}
\eeq
The world-volume action for the translational modulus is obtained from our calculation as
\beq
{\Delta}S_{\rm trans}=\frac{T}{2}\int dt\,dx\,dy \,
\left(\partial_{p}z_{0}\partial^{p}z_{0}\right),
\label{wsactionforboth}
\eeq
as it should be on general grounds. Numerically (under our choice of parameters) the total tension is $$T= 0.998 m_\varphi^3\,,$$ the integral over the $\varphi$ derivative term is $$T_{\varphi} =0.844 m_\varphi^3\,,$$ and the integral over the $\chi$ derivative term is $$T_\chi = 0.154 m_\varphi^3\,.$$

\section{Adding a spin-orbit interaction}

Previously translational and rotational modes were independent. Now let us introduce an additional term for spin-orbit interaction,
\beq
{\cal L}_{so}=-\varepsilon(\partial_i\chi^i)^2\,.
\eeq
The {\em raison d'etre} of such a term and its possible origin is discussed in \cite{NShV}.

To the leading order in $\varepsilon$, now we obtain an additional term in the action,
\begin{eqnarray}
\Delta S_{3}&=&-\varepsilon\chi_{0}^{2}m_{\varphi}I_{2}\int dtdxdy
\left[(\partial_{k}z_{0})(\partial_{l}z_{0})S^{k}S^{l}+S^{3}S^{3}+2(\partial_{k}z_{0})S^{k}S^{3}\right]
\nonumber\\[3mm]
&-&
\varepsilon\frac{\chi_{0}^{2}}{m_{\varphi}}I_{1}\int dtdxdy\left(\partial_{k}S^{k}\right)^{2},\qquad k=1,2.
\end{eqnarray}
We see that the structure of this entangled term is quite different from that in \cite{MShYu} for 
the Abrikosov-Nielsen-Olesen string. 
The entanglement between the translational and orientational moduli of a special form is evident.

In higher orders in $\varepsilon$ the spin-orbit interaction will also change the formula for the   $\chi(z)$
solution compared to that following from Eq. (12).

\section{Conclusions}

We considered the simplest model with a  domain wall and non-Abelian moduli fields, localized on it. 
Instability of the solution without the additional $\chi$ field was shown, 
and then the numerical solution with $\chi\neq 0$ was found. 
The latter implies the existence of extra rotational moduli on the wall world volume. 
Then we considered this model with addition of a spin-orbit interaction.

\section*{Acknowledgments}

The work of M.S. is supported in part by DOE grant DE-SC0011842.
Kind hospitality and support extended to M.S. during his stay at IHES is acknowledged.

\end{document}